\documentclass{svjour3}
\journalname{Theoretical Chemistry Accounts}
\usepackage{dcolumn}

\usepackage{amsmath}
\usepackage{braket}
\usepackage{hyperref}
\renewcommand{\vec}[1]{{\rm\bf #1}}
\DeclareMathOperator{\erf}{\mathrm{erf}}
\newcommand{\de}{\mathrm{d}}

\begin{document}

\title{Atomic effective potentials for starting molecular electronic structure calculations }

\author{Dimitri N. Laikov \and Ksenia R. Briling}

\institute{
 D. N. Laikov \at
 Chemistry Department, Moscow State University, 119991 Moscow, Russia \\
\email{laikov@rad.chem.msu.ru}
\and
 K. R. Briling \at
 Chemistry Department, Moscow State University, 119991 Moscow, Russia \\
\email{briling@rad.chem.msu.ru}
}

\maketitle

\begin{abstract}
Atomic effective one-electron potentials
in a compact analytic form in terms of
a few Gaussian charge distributions
are developed, for Hydrogen through Nobelium,
for starting molecular electronic structure calculations
by a simple diagonalization.
For each element, all terms but one
are optimized in an isolated-atom Hartree--Fock calculation,
and the last one is parametrized on a set of molecules.
This one-parameter-per-atom model
gives a good starting guess for typical molecules
and may be of interest even on its own.
\end{abstract}

\section{Introduction}

Every molecular electronic structure calculation
needs a good starting guess.
Most likely the earliest and still widely used
extended H\"uckel~\cite{H1963} theory gives
a minimal-basis solution that can be then projected~\cite{KSKWP1967}
onto the working basis set.
The underlying minimal basis should be given in some way ---
some workers~\cite{SBBEGJKMNSWDM1993} used
short three-Gaussians-per-shell sets~\cite{HAKRST1984}
for atoms up to Radon, with known shortcomings.
We thought it would be handy to have a tool to build
a better minimal set straight from the working set ---
clearly, atomic Hartree--Fock calculations could be done for that,
but even then a starting guess is needed.
We have found that the optimized effective potentials~\cite{SH1953,TS1976}
for atoms can be given, accurately enough, in a very compact form
in terms of Gaussian charge distributions ---
they can be pre-tabulated and used to get the starting guess for an atom
with one diagonalization of an easily-computable matrix.
One may wonder whether this has already been done for all elements,
but we see only works of this kind~\cite{MSBG2011}
with rather complicated, and even piecewise,
analytic form of the potential.

Another well-known starting guess for molecules
uses an effective one-electron Hamiltonian
of density-functional theory~\cite{KS1965}
within the approximation~\cite{H1985}
of overlapped but frozen atomic densities.
The nonlinearity of the exchange-correlation functionals
makes the matrix elements of the potential
tractable only by numerical integration,
for which good schemes~\cite{B1988,LKO2018} are known
but need to be implemented.
A linearized version thereof is simply the sum of
fixed neutral-atom potentials taken from
a density-functional calculation,
which, remarkably, has been introduced and benchmarked
\textit{for the first time}
only a few months ago~\cite{Lehtola2019}.
There is also a Hartree--Fock analog~\cite{LZDG2006}
that needs up to three-center two-electron integrals,
or its spherical approximation~\cite{AC2001}
with less costly models of the direct and exchange terms.

Of course, the idea
to approximate the molecular effective one-electron potential
by a sum of atom-centered functions
is not new and dates back to the early density-functional works~\cite{SF1975},
where the Coulomb part was calculated using density fitting
and the exchange-correlation potential was also fitted,
either on the grid~\cite{DCS1979}, or,
in the special case of Hartree--Fock--Slater~\cite{S1951} theory,
by an elegant grid-free variational method~\cite{D1986}
with analytical evaluation of all three-center integrals.
Though the approximation of both Coulomb and exchange-correlation terms
by an electrostatic potential arising from a sum of atom-centered (Gaussian-type)
effective charge distributions was put forward by one of us~\cite{L1997}
more than 20 years ago, there seems to be a new interest
in using this functional form to derive transferable atomic potentials
for molecules in educational work~\cite{NW2017} and beyond~\cite{W2019,W2019b}.

Our atom-optimized compact representations
can also be adopted and adapted for molecules
by ``capping'' each atom with one more adjustable potential term
to cancel the $-1/r$ asymptotic behavior and make it
short-ranged, this way we get a new tunable model
worth being studied.
Our tests show it to be at least as good
as an extended H\"uckel guess,
we also compare it with the sum of tailless
Hartree--Fock--Slater-based neutral-atom potentials
and we see that our tunable model often works much better,
so we report it here in the hope
that it may be helpful for others.

\section{Theory}

For an atom, our effective potential $v(\vec r)$ is the sum
of its bare-nuclear potential $v_{\mathrm{n}}(\vec r)$
and an approximate (direct and exchange) screening term
\begin{equation}
\label{eq:v}
v_{\mathrm{a}}(\vec r) =
\sum\limits_{i=1}^{n} c_i \, \frac {\erf \big( \sqrt{a_i} \, |\vec r| \big)}{ |\vec r| }
\end{equation}
in the form of Coulomb potential of Gaussian~\cite{B1950} charge distributions.
We optimize the exponents $a_i$ and coefficients $c_i$,
with the constraint
\begin{equation}
\label{eq:q}
\sum\limits_{i=1}^{n} c_i = Q - 1,
\end{equation}
$Q$ being the nuclear charge,
to minimize the measure
\begin{equation}
\label{eq:e}
\mathcal{E} = \tfrac12 \iint \frac{
  \big(\rho_v (\vec r_1) - \rho_0(\vec r_1) \big)
  \big(\rho_v (\vec r_2) - \rho_0(\vec r_2) \big)
  }{\left| \vec r_1 - \vec r_2 \right|}
  \,\de^3 \vec r_1
  \,\de^3 \vec r_2
\end{equation}
of the difference
between the Hartree--Fock $\rho_0$
and the effective-potential-based $\rho_v$ densities.
Spherically-averaged configurations~\cite{L2019b}
are taken for atoms with open-shell ground states.
Eq.~(\ref{eq:q})
ensures the right asymptotic behavior
of $v(\vec r)$ as $|\vec r| \to \infty$.
We use the four-component scalar-relativistic~\cite{D1994} Hamiltonian
and a finite Gaussian~\cite{VD1997} nucleus model
for all atoms Hydrogen through Nobelium.

Table~\ref{tab:at} shows the number of terms $n$
and the value of $\mathcal E$ for the groups of atoms,
the overall accuracy is quite good, $\mathcal E < 0.0004$;
$n-1$ terms would raise $\mathcal E$ by about 10 times,
while $n+1$ can hardly do any better
and would often lead to run-away solutions.
We also check the atomic Hartree--Fock energy errors $\Delta E$
within a two-component scalar-relativistic approximation~\cite{L2019a}
thanks to its bounded\-ness from below.

\begin{table}
\caption{\label{tab:at}Atomic parameters.}
\begin{tabular}{lrrcc}
\hline\noalign{\smallskip}
Atoms  & $n$ & $a_{n+1}$  & $\mathcal E$\textsuperscript{a} & $\Delta E$\textsuperscript{a,b} \\
\noalign{\smallskip}\hline\noalign{\smallskip}
H      &  0  & $1/3$      &          &        \\
He     &  1  & $1/3$      & 0.000024 & 0.0004 \\
Li     &  1  & $1/16$     & 0.000154 & 0.0050 \\
Be     &  2  & $1/16$     & 0.000075 & 0.0040 \\
B--Ne  &  2  & $1/3$      & 0.000320 & 0.0094 \\
Na     &  2  & $1/32$     & 0.000379 & 0.0135 \\
Mg     &  3  & $1/32$     & 0.000099 & 0.0069 \\
Al--Ar &  3  & $1/8$      & 0.000143 & 0.0103 \\
K--Ca  &  3  & $1/32$     & 0.000275 & 0.0138 \\
Sc--Zn &  4  & $1/6$      & 0.000029 & 0.0191 \\
Ga--Kr &  4  & $1/12$     & 0.000180 & 0.0210 \\
Rb--Sr &  4  & $1/32$     & 0.000085 & 0.0159 \\
Y--Cd  &  4  & $1/8$      & 0.000391 & 0.0326 \\
In--Xe &  5  & $1/12$     & 0.000035 & 0.0225 \\
Cs--Yb &  5  & $1/32$     & 0.000205 & 0.0519 \\
Lu--Hg &  5  & $1/12$     & 0.000239 & 0.0537 \\
Tl--Rn &  6  & $1/12$     & 0.000069 & 0.0450 \\
Fr--No &  6  & $1/32$     & 0.000297 & 0.0835 \\
\noalign{\smallskip}\hline
\end{tabular}
\begin{flushleft}
\textsuperscript{a} The largest value in the row.\\
\textsuperscript{b} Energy error.
\end{flushleft}
\end{table}

For a molecule with $N$ atoms, the atomic potential tails of $-1/r$
would have summed up to an unphysical $-N/r$;
as a lesser evil, we decided to ``cap'' each atom
with one more term having $c_{n+1} = 1$ and a to-be-optimized $a_{n+1}$.
(Setting $c_{n+1} = (N-1)/N$ may be kept as an option.)
This also gives us the freedom to adjust the potential well depth
in the valence region to mimic a typical molecular environment.

For the optimization, we tried two objective functions:
(a) the energy-based
\begin{equation}
f_\mathrm{E} = \sum_i \left( \braket{ \phi_i | \hat F^0 | \phi_i } - \braket{ \phi_i^0 | \hat F^0 | \phi_i^0 } \right),
\end{equation}
or (b) the overlap-based
\begin{equation}
f_\mathrm{S} = \sum_j \left( 1 - \sum_i \left| \braket{ \phi_i | \phi_j^0 } \right|^2 \right),
\end{equation}
where the wavefunctions $\phi_j^0$
and the Fock operator $\hat F^0$ are
taken from the reference self-consistent field calculation,
whereas $\phi_i$ are our approximate solutions,
the (spin and space) labels $i$ and $j$ run over the occupied valence set.
The functions, either $f_\mathrm{E}$ or $f_\mathrm{S}$,
are to be summed up over a training set of molecules
and minimized with respect to the atomic parameters.

Our thus derived potentials can work equally well
with any all-electron formalism, both two- and four-component.
They can also be easily adapted for use with
effective core potentials~\cite{KBT76} by subtracting
the core charge $Q_0$ from $c_i$ referring to the largest $a_i$,
\begin{equation}
\bar{c}_i =
\left\{
\begin{array}{cl}
0, & i<\bar{n}, \\
\sum_{j=1}^{\bar{n}} c_j - Q_0,  & i=\bar{n}, \\
c_i, & i>\bar{n},
\end{array}
\right.
\end{equation}
with $\bar{n}$ such that $0\le \bar{c}_i \le c_i$,
where $a_i > a_{i+1}$ for all $i$.

\section{Calculations}

The problem of Eq.~(\ref{eq:e}) was solved
by a code written by one of us (DNL)
within the atomic structure program~\cite{L2019b},
while the further optimization of parameters on molecules
was done by an original program\footnote{To be found at \url{https://github.com/briling/aepm}.} written by the other (KRB),
and then the starting guess for molecules
was double-checked by each of us \textit{independently}
by one's own molecular electronic structure code.

The Hartree--Fock-optimized atomic parameters
for Helium through Nobelium
are tabulated in the Supplementary material.
The reference molecular calculations were done
with the PBE~\cite{PBE1996} density functional,
the scalar-relativistic approximation~\cite{L2019a},
and the L1 basis~\cite{L2005,L2019b}.

First, we optimized the $a_{n+1}$ parameters for atoms H through Cl (without He and Ne)
on the neutral closed-shell subset of molecules~\cite{L2011}.
For H and B--F we got the values strikingly close to $1/3$ ---
this, together with the flatness near the minimum,
led us to have all parameter values within groups of atoms
constrained to be equal during the optimization and
rounded afterwards to simple fractions.
For the heavier elements, skipping Pr--Tm and Np--Md with their open-shell states,
we have built up a smaller set of prototypical molecules
and gotten the rest of values shown in Table~\ref{tab:at}.
On average, we get the energy error \textit{per atom} $f_E /N \approx 0.005$ au
for the lighter, and $\approx 0.009$ au for the heavier, subsets.

Taking atomic Hartree--Fock--Slater densities as $\rho_0$ in Eq.~(\ref{eq:e})
and setting the constraint with $Q$ on the right hand side of Eq.~(\ref{eq:q}),
we have optimized another set of potentials of Eq.~(\ref{eq:v})
now with $n$ simply equal to the highest occupied principal quantum number
in the atom, they are also tabulated in the Supplementary material
for all 102 elements.
Symptomatically, we get $a_1 \approx 0.08$ for Hydrogen,
rather diffuse compared to our optimized value of $1/3$.
On our set of molecules,
we get systematically higher error measures $f_E$ in most cases
when the parametrized potentials are replaced
with their Hartree--Fock--Slater-based analogs,
typically by $\sim 10$ times for the organic subset
(a full listing can be found in the Supplementary material),
thus our parametrization is indeed helpful.

\section{Conclusions}

It would be straightforward to implement our starting guess,
as an option, into any electronic structure code ---
for Gaussian basis sets, all the integrals
can be easily computed analytically~\cite{B1950,MD1978}.
On the other hand, this elegant one-parameter-per-atom model
may be of interest even on its own.
Followed by an energy correction in the spirit
of second-order perturbation theory~\cite{HK1983,WP2003,DGG2009,DGG2010},
it may try to replace the iterative self-consistent solution
in lower-accuracy applications.

An unnamed reader has warned us of the dangers of black boxes~\cite{MBP2009},
and though we find our potentials to work well in the raised case of ferrocene,
one should always keep in mind that a guess is a guess,
however good it may be on average.

\section{Supplementary material}

File format description (\texttt{readme.txt}),
atomic parameters based on Hartree--Fock (\texttt{ac.txt})
and Hartree--Fock--Slater (\texttt{ac0.txt}) theory,
and error measures for all molecules (\texttt{fm.txt}).

\clearpage

\textit{file ``readme.txt''}
\begin{verbatim}
Supplementary material
for
"Atomic effective potentials for starting molecular electronic structure calculations"
by
Dimitri N. Laikov and Ksenia R. Briling


The file "ac.txt" holds parameters of atomic effective potentials
for atoms Helium through Nobelium in the format:
the atomic number and the number of terms n
followed by n pairs of the exponent and coefficient, one per line.

For example:

  18 3
+.70097817629160849215e+02  +.33299240013050428956e+01
+.42193314636035713068e+01  +.77190281538706445101e+01
+.35198503878294074576e+00  +.59510478448243125943e+01

for atomic number 18 (Argon) 3 (three) terms are listed
with exponents a1 = 70.1, a2 = 4.2, a3 = 0.35,
and coefficients c1 = 3.3, c2 = 7.7, c3 = 6.0 (rounded).

The Hartree-Fock-Slater analogs are in the file "ac0.txt",
for comparison.

Molecular tests are documented in the file "fm.txt":
for each molecule (column 1), the energy (Eq. 4) errors
for parametrized (column 2) and Hartree-Fock-Slater-based (column 3)
potentials are listed followed by the overlap (Eq. 5)
errors (columns 4 and 5), all values in a.u.

\end{verbatim}

\newpage

\textit{file ``ac.txt''}
\begin{verbatim}
   2 1
+.18865345899608519089e+01  +.10000000000000000000e+01
   3 1
+.19854870701524917779e+01  +.20000000000000000000e+01
   4 2
+.47445861849777778539e+01  +.16757452423122399273e+01
+.27924701370840662020e+00  +.13242547576877600727e+01
   5 2
+.60338581393756149699e+01  +.21592228097073424342e+01
+.22966528454630481676e+00  +.18407771902926575658e+01
   6 2
+.83684238262991903102e+01  +.23490566496000995778e+01
+.31758238510185922825e+00  +.26509433503999004222e+01
   7 2
+.10933999496275620559e+02  +.25402896286376577300e+01
+.43457823405570917314e+00  +.34597103713623422700e+01
   8 2
+.13822779569568998053e+02  +.26700915408359251462e+01
+.61638076315423918696e+00  +.43299084591640748538e+01
   9 2
+.16696221288447185150e+02  +.28413526537091200589e+01
+.80696743351842942568e+00  +.51586473462908799411e+01
  10 2
+.19447665246333681427e+02  +.30482912873644855001e+01
+.10081157441421303946e+01  +.59517087126355144999e+01
  11 2
+.22043514485429395037e+02  +.33181662867767014502e+01
+.10688208368282481248e+01  +.66818337132232985498e+01
  12 3
+.35680895797762356478e+02  +.25473056612272425785e+01
+.29023990296953043232e+01  +.49309783934737768009e+01
+.39191184585485700913e+00  +.35217159452989806206e+01
  13 3
+.34328377368288002050e+02  +.32064176805161995937e+01
+.18953919764518969282e+01  +.69511251543690147803e+01
+.12243916188522635365e+00  +.18424571651147856260e+01
  14 3
+.40176352944236500295e+02  +.32525044721783944930e+01
+.22394952559801087748e+01  +.72406048949908103012e+01
+.13204220229571035912e+00  +.25068906328307952058e+01
  15 3
+.46664937337468767954e+02  +.32863376244960580497e+01
+.26279568276824812544e+01  +.74809093146721600666e+01
+.15940360302607907978e+00  +.32327530608317818836e+01
  16 3
+.54215297785332151257e+02  +.32836304060875017809e+01
+.31676473151453729976e+01  +.75042023824894442235e+01
+.22671769463490918175e+00  +.42121672114230539956e+01
  17 3
+.62030532593708839335e+02  +.32996554318756391397e+01
+.37003973360077539976e+01  +.75888303224369905897e+01
+.28974576291563425825e+00  +.51115142456873702706e+01
  18 3
+.70097817629160849215e+02  +.33299240013050428956e+01
+.42193314636035713068e+01  +.77190281538706445101e+01
+.35198503878294074576e+00  +.59510478448243125943e+01
  19 3
+.81433339114890036429e+02  +.32644630721960025775e+01
+.49621419831026952504e+01  +.79201946092260660689e+01
+.37318942904625840592e+00  +.68153423185779313535e+01
  20 3
+.74415699626934158651e+02  +.38596365529694548520e+01
+.40715111934796436426e+01  +.86137740964443135681e+01
+.28671657971452114743e+00  +.65265893505862315799e+01
  21 4
+.10120791061796357900e+03  +.32849848247331625325e+01
+.64467237284313891045e+01  +.78787365837762881328e+01
+.78340450650304679730e+00  +.48448234863806428085e+01
+.22063826662337555876e+00  +.39914551051099065262e+01
  22 4
+.11439209603243822314e+03  +.32219238174458418393e+01
+.77510310640834962077e+01  +.74374376218330072798e+01
+.13409079006639176087e+01  +.45006797826816257199e+01
+.29105515283701394607e+00  +.58399587780395251610e+01
  23 4
+.12816680647954118893e+03  +.31682054760600673936e+01
+.92166956666162262490e+01  +.69977642782110986564e+01
+.18583538531410260272e+01  +.50682125549198122790e+01
+.34033218181509780281e+00  +.67658176908090216710e+01
  24 4
+.14277227533657647044e+03  +.31176645625448774262e+01
+.10841336380933656177e+02  +.66176777414958036611e+01
+.23247918847354291701e+01  +.58188197729153805409e+01
+.38605354233477900560e+00  +.74458379230439383718e+01
  25 4
+.15833292213000518305e+03  +.30681465399637835553e+01
+.12618932464144514378e+02  +.63132925341478953756e+01
+.27513177146687477771e+01  +.65983971904932756772e+01
+.43139865242691528668e+00  +.80201637353950453919e+01
  26 4
+.17159971217722957215e+03  +.30934781740775755236e+01
+.12848889315993880055e+02  +.70267535234960223497e+01
+.24208579434027236201e+01  +.72576961406448163284e+01
+.44179048296062721916e+00  +.76220721617815857983e+01
  27 4
+.18674822291999738102e+03  +.30895907193295115615e+01
+.13743136771816508454e+02  +.73157772470042668424e+01
+.23981274633981865605e+01  +.81771064014486378193e+01
+.45853074139738459791e+00  +.74175256322175837767e+01
  28 4
+.20317863825379029239e+03  +.30750315752041584293e+01
+.14904507246734576915e+02  +.74575722713464167488e+01
+.25051824055969202871e+01  +.90765297154785403004e+01
+.48217231383091866431e+00  +.73908664379708845215e+01
  29 4
+.22069456355904439006e+03  +.30558773005594047794e+01
+.16228087185242245482e+02  +.75352810704668349367e+01
+.26799011471909331258e+01  +.99201387162312768935e+01
+.51163451896920490175e+00  +.74887029127424833903e+01
  30 4
+.23922585729085248403e+03  +.30345836308340147209e+01
+.17676284456609932035e+02  +.75805648880574860050e+01
+.28967970112707354043e+01  +.10716681306345864678e+02
+.54578412163007880632e+00  +.76681701747626345966e+01
  31 4
+.20777682210003922500e+03  +.36665589239870065015e+01
+.11856343239908128003e+02  +.10755736410261744995e+02
+.14354572368215609153e+01  +.13141530303552718351e+02
+.17389779129425224574e+00  +.24361743621985301523e+01
  32 4
+.22524685392781287214e+03  +.36359325279082192775e+01
+.12852764572486531416e+02  +.10984641903702068363e+02
+.15215113800917213721e+01  +.13803155127032158374e+02
+.13230147772140648943e+00  +.25762704413575539853e+01
  33 4
+.24790861102527441995e+03  +.35589986946919740276e+01
+.14300295034709391484e+02  +.11010431477805105621e+02
+.16861444954415892161e+01  +.14324814816160803325e+02
+.13416499137137873264e+00  +.31057550113421170261e+01
  34 4
+.27386499439661639875e+03  +.34662210340132432196e+01
+.16081100126028409415e+02  +.10903636951521558137e+02
+.19320318837422158703e+01  +.14582668431270511624e+02
+.17923310561826160870e+00  +.40474735831946870188e+01
  35 4
+.30001299028853977035e+03  +.33921113983124644045e+01
+.17856104656403406708e+02  +.10850224593864627235e+02
+.21685919518796942007e+01  +.14848339655969876824e+02
+.21670708306814742347e+00  +.49093243518530315364e+01
  36 4
+.32669023865557319639e+03  +.33305063007909982237e+01
+.19647882619248066884e+02  +.10831107022676644388e+02
+.23989880465652471323e+01  +.15121828835488210045e+02
+.25103028444531387577e+00  +.57165578410441473425e+01
  37 4
+.37163737826238113430e+03  +.31466666484776166178e+01
+.23248544533094692908e+02  +.10341351681057330980e+02
+.28726598551454175825e+01  +.15563074945782764186e+02
+.27773911546919278777e+00  +.69489067246822882167e+01
  38 4
+.37072341161894166681e+03  +.33133131598951131789e+01
+.22369820917706722631e+02  +.11069248049555791375e+02
+.27431178263461401512e+01  +.15751396571059824070e+02
+.22670958449869717963e+00  +.68660422194892713765e+01
  39 4
+.38194339180832728613e+03  +.33897687777580409061e+01
+.22657005971801517501e+02  +.11515238880840111505e+02
+.27818156103527597502e+01  +.15744363112944809796e+02
+.23142082707081156801e+00  +.73506292284570377920e+01
  40 4
+.40412975489707895010e+03  +.33912387012959417039e+01
+.23913120490512053721e+02  +.11698493461811265601e+02
+.29281223941744981287e+01  +.15897124348300313556e+02
+.25096601775828317782e+00  +.80131434885924791388e+01
  41 4
+.43200075322658102319e+03  +.33619764754289541514e+01
+.25649609677505364113e+02  +.11759141278005034150e+02
+.31272636914975559143e+01  +.16141877715462901120e+02
+.27614878489011475330e+00  +.87370045311031105790e+01
  42 4
+.46268918558642525532e+03  +.33241931072269496919e+01
+.27620091978879956212e+02  +.11775004054054392252e+02
+.33502415418868519603e+01  +.16421219554119644948e+02
+.30463722353499470739e+00  +.94795832845990131073e+01
  43 4
+.49504830506721169066e+03  +.32859988207200458037e+01
+.29724523813339572593e+02  +.11777513760113202247e+02
+.35840719287190991983e+01  +.16713996790559294240e+02
+.33542029325897034058e+00  +.10222490628607457709e+02
  44 4
+.50788917516825208035e+03  +.33568004375740026291e+01
+.30119638007448954451e+02  +.12183651199011928789e+02
+.35989334168125757504e+01  +.16732947240496758997e+02
+.36368976141299972161e+00  +.10726601122917309585e+02
  45 4
+.52496383750584292732e+03  +.34042696795165551855e+01
+.30919224799048707681e+02  +.12498923939454255853e+02
+.36440391113842645119e+01  +.16870466086010282043e+02
+.39297773558866046385e+00  +.11226340295018906918e+02
  46 4
+.54537148473898028396e+03  +.34352063845776484904e+01
+.32017351723397207779e+02  +.12743509881914651382e+02
+.37145429669883926110e+01  +.17096800752115824492e+02
+.42339611975503888333e+00  +.11724482981391875635e+02
  47 4
+.56848334028080858164e+03  +.34544421984266276586e+01
+.33345924933420031056e+02  +.12934294037833886018e+02
+.38059246866006447837e+01  +.17389101155187288809e+02
+.45499468116272807546e+00  +.12222162608552197514e+02
  48 4
+.59388588982128796181e+03  +.34652635118576361543e+01
+.34862597588803264317e+02  +.13083885898252915897e+02
+.39146777891016737707e+01  +.17730959451884514142e+02
+.48779609479059794305e+00  +.12719891138004933806e+02
  49 5
+.74504533341674428055e+03  +.29874634939188884888e+01
+.47589716263493248893e+02  +.11154282804456501447e+02
+.59712038282804846794e+01  +.17314739405885692879e+02
+.74300422091938889735e+00  +.14971163078231132275e+02
+.84660995560378415338e-01  +.15723512175077849100e+01
  50 5
+.79702852654173314139e+03  +.29316341933211216742e+01
+.51533843511592459550e+02  +.10995396237542593334e+02
+.65092596715707926475e+01  +.17451448107477579837e+02
+.80544055332868458826e+00  +.15716102363338380092e+02
+.66326146535977302557e-01  +.19054190983203250623e+01
  51 5
+.85433263573360495731e+03  +.28687229851761811778e+01
+.56030106441107747302e+02  +.10796297417605338766e+02
+.71434023126589638652e+01  +.17532917379237792734e+02
+.88347126319429414394e+00  +.16400893313073775167e+02
+.67335798553194526531e-01  +.24011689049069121549e+01
  52 5
+.91216680013767399628e+03  +.28147490944765846995e+01
+.60593809682671006622e+02  +.10621553493355722879e+02
+.78128697037188814937e+01  +.17535679854859787483e+02
+.98647849331351539671e+00  +.16749961304308713660e+02
+.10139508056351640081e+00  +.32780562529991912778e+01
  53 5
+.97123265740465906255e+03  +.27669360798826062817e+01
+.65252543199773385315e+02  +.10468708135547245616e+02
+.84954736819143606181e+01  +.17546759842438655658e+02
+.10874117592443088735e+01  +.17126418429611765221e+02
+.12699302114117239117e+00  +.40911775125197272236e+01
  54 5
+.10318412270302938313e+04  +.27237667219329389421e+01
+.70039902307632663117e+02  +.10330894101872670405e+02
+.91969976448947273582e+01  +.17563256036021080262e+02
+.11869856131289064870e+01  +.17530860453395113778e+02
+.14857777514764354720e+00  +.48512226867781966123e+01
  55 5
+.11493826345739148054e+04  +.25533576720595384602e+01
+.82074003023619112491e+02  +.95665281654822195899e+01
+.11211986732106572843e+02  +.17061280408746025814e+02
+.15279462333396209482e+01  +.17732211210674003336e+02
+.20092255135604541527e+00  +.70866225430382127994e+01
  56 5
+.11696947757016588526e+04  +.26213312447246266653e+01
+.81771576797198378601e+02  +.99171262225801874566e+01
+.11074931503288440006e+02  +.17308600695796790032e+02
+.15039225873956188253e+01  +.18136517413864287698e+02
+.17008617153559455691e+00  +.70164244230341081479e+01
  57 5
+.12222841796269503170e+04  +.26195561375108166129e+01
+.85307385385270565408e+02  +.99577548957140074202e+01
+.11534893922454100338e+02  +.17452839372282196061e+02
+.15652164957866740791e+01  +.18426529369428939588e+02
+.17220089046882704072e+00  +.75433202250640403172e+01
  58 5
+.12720807279952528946e+04  +.26276603866922210186e+01
+.88366575508642884658e+02  +.10036579425672154421e+02
+.11818291666733136329e+02  +.17732191855663436105e+02
+.16071799883505355389e+01  +.18927635504515411216e+02
+.18079491474660023699e+00  +.76759328274567772394e+01
  59 5
+.13221469857952937682e+04  +.26380802907857773228e+01
+.91396564321374118052e+02  +.10114040279413738988e+02
+.12107117079722300083e+02  +.17993264035731876427e+02
+.16537813057728806520e+01  +.19493835869390085116e+02
+.18837281505285845328e+00  +.77607795246785221458e+01
  60 5
+.13725263581315083456e+04  +.26506050159872611378e+01
+.94382563397985009911e+02  +.10193310028445527863e+02
+.12388920046052041834e+02  +.18254727161343399934e+02
+.17024088534425037373e+01  +.20085792564483675314e+02
+.19527976756889851334e+00  +.78155652297401357514e+01
  61 5
+.14232939068289442976e+04  +.26649446538031491736e+01
+.97328657834613684470e+02  +.10274459071687807490e+02
+.12661617245227912706e+02  +.18522471669075388246e+02
+.17523980768357151358e+01  +.20688422677958256615e+02
+.20168624380514636287e+00  +.78497019274753984755e+01
  62 5
+.14745487132299602158e+04  +.26807923239534211711e+01
+.10024430189239065721e+03  +.10356881026567634681e+02
+.12925522095087921053e+02  +.18799417404521530458e+02
+.18034727018822441493e+01  +.21293977241942228382e+02
+.20769358126601012487e+00  +.78689320030151853081e+01
  63 5
+.15264150536618301112e+04  +.26978640905721766211e+01
+.10314004849790670614e+03  +.10439868003977639808e+02
+.13181653182291660523e+02  +.19087352645971370073e+02
+.18554646230381475835e+01  +.21897912345772990911e+02
+.21336744705076069654e+00  +.78770029137058225877e+01
  64 5
+.15789784735126056335e+04  +.27159305699306148774e+01
+.10602501158566269475e+03  +.10522796214675617561e+02
+.13431158961640053435e+02  +.19387452129770386514e+02
+.19082409938489719327e+01  +.22497378964585917405e+02
+.21875121147602400890e+00  +.78764421210374636418e+01
  65 5
+.16279257815186742923e+04  +.27411226932956958183e+01
+.10858046918725347295e+03  +.10616350835073855907e+02
+.13712227614246241933e+02  +.19538239265378358275e+02
+.19934065152409421501e+01  +.23132099102842867011e+02
+.22607538133284423461e+00  +.79721881034092229887e+01
  66 5
+.16757366368652999276e+04  +.27697835225437479078e+01
+.11092891605390537645e+03  +.10722503380066446967e+02
+.13944492993755349266e+02  +.19736802312959688194e+02
+.20713613685444629290e+01  +.23730486350905376132e+02
+.23293090457243363435e+00  +.80404244335247407995e+01
  67 5
+.17224510232360898468e+04  +.28015830284582851640e+01
+.11309508331285164387e+03  +.10838293069461611601e+02
+.14135448633292121789e+02  +.19978990892678130435e+02
+.21434468833925621800e+01  +.24293655068020399645e+02
+.23939934892582106229e+00  +.80874779413815731556e+01
  68 5
+.17681273694341087398e+04  +.28361728534561255897e+01
+.11510451705123839658e+03  +.10961108480750576628e+02
+.14292436046614584054e+02  +.20261631825154297223e+02
+.22106447962734520737e+01  +.24823569072814496512e+02
+.24552713219420616199e+00  +.81175177678245040477e+01
  69 5
+.18128387097425374031e+04  +.28732306358723144930e+01
+.11698049742273002375e+03  +.11088803657353719962e+02
+.14422182620212654739e+02  +.20582107554913643492e+02
+.22736936308261366496e+01  +.25322405700084024151e+02
+.25134177827762767419e+00  +.81334524517762979017e+01
  70 5
+.18566685254316594285e+04  +.29124652371565373783e+01
+.11874345107133860150e+03  +.11219670947580419406e+02
+.14530686065898029811e+02  +.20938069255347597683e+02
+.23331738881401942761e+01  +.25792350278740494423e+02
+.25686096532697333606e+00  +.81374442811749511096e+01
  71 5
+.18997375959050908652e+04  +.29535955441201374457e+01
+.12041203401518791747e+03  +.11352349978687230449e+02
+.14623290920474742214e+02  +.21327243843421123933e+02
+.23895731400544327792e+01  +.26235577299844828812e+02
+.26209875138329045085e+00  +.81312333339266793605e+01
  72 5
+.19673161513254125348e+04  +.29621161002957168217e+01
+.12434022381142522564e+03  +.11361133113996146592e+02
+.15073083794221049984e+02  +.21578072665372014527e+02
+.24808185704179228484e+01  +.26565925422554728476e+02
+.26617131686220868124e+00  +.85327526977813935829e+01
  73 5
+.20700813765288223264e+04  +.29277615621936452850e+01
+.13149175287923276758e+03  +.11205512222519954671e+02
+.15991001325343890074e+02  +.21617220829977906263e+02
+.26318773077354023292e+01  +.27084534015420166982e+02
+.28043101980165810474e+00  +.91649713698883267988e+01
  74 5
+.21895493581497282395e+04  +.28789204905262306730e+01
+.14025127395857804340e+03  +.10991065963882566690e+02
+.17133447569749620252e+02  +.21575368819978679375e+02
+.28098978707573803615e+01  +.27659569431312725045e+02
+.29967949093936527104e+00  +.98950752942997982170e+01
  75 5
+.23184462036806926004e+04  +.28264828916719967410e+01
+.14999379763687080270e+03  +.10759035990377656099e+02
+.18404731677310796045e+02  +.21506658786388130323e+02
+.30016080614767283756e+01  +.28237252265691354114e+02
+.32187252568656625317e+00  +.10670570065870862723e+02
  76 5
+.23880490707156892183e+04  +.28473877823315644884e+01
+.15377349232258009740e+03  +.10796529419101561500e+02
+.18857909721345863635e+02  +.21734431161786750354e+02
+.31100863312692503356e+01  +.28306314130597089890e+02
+.34445180147113784846e+00  +.11315337506183033768e+02
  77 5
+.24674297047382197439e+04  +.28595480701182473255e+01
+.15846189701303072718e+03  +.10796791135751348468e+02
+.19439726509029725641e+02  +.21922974276466809049e+02
+.32301679883724033225e+01  +.28429995460609494969e+02
+.36862565727337615862e+00  +.11990691057054100188e+02
  78 5
+.25543592830018635085e+04  +.28662462001737870994e+01
+.16379362965973422753e+03  +.10774465878440480825e+02
+.20109681026841000826e+02  +.22086385323186230952e+02
+.33577853513983553563e+01  +.28589133015470682202e+02
+.39407497043736374318e+00  +.12683769582728818922e+02
  79 5
+.26479423678868878637e+04  +.28690702858987777134e+01
+.16965288496596832204e+03  +.10736842366501742438e+02
+.20849001035167128167e+02  +.22231851688015446299e+02
+.34909991955041559319e+01  +.28774342144019749205e+02
+.42063212480622336363e+00  +.13387893515564284345e+02
  80 5
+.27479281346759772134e+04  +.28687766249959008053e+01
+.17599878082315510778e+03  +.10687584444214812912e+02
+.21649674830696111463e+02  +.22362991007380984298e+02
+.36289678032455827459e+01  +.28981078296242676723e+02
+.44820670782959366692e+00  +.14099569627165625262e+02
  81 6
+.31644430017902674779e+04  +.25898355236158902471e+01
+.21572530656955273349e+03  +.96246431246896855497e+01
+.27161342191636469259e+02  +.20809716003886912456e+02
+.44914277801281687709e+01  +.29952533588261982205e+02
+.60044016507879410021e+00  +.15700427135109986694e+02
+.68530932050186837842e-01  +.13228446244355428485e+01
  82 6
+.33326463110290519887e+04  +.25510270843967342072e+01
+.22900310905632420178e+03  +.94477355440436975334e+01
+.28975969438335592188e+02  +.20605202222691935705e+02
+.48105585257086867361e+01  +.30226868107463972659e+02
+.64280898159203082927e+00  +.16563554283376984155e+02
+.47924330717300866072e-01  +.16056127580266757405e+01
  83 6
+.35122165088553368970e+04  +.25107682966586122455e+01
+.24339214593688772585e+03  +.92637277362678042288e+01
+.30978532465816319835e+02  +.20354334911567376745e+02
+.51724235894520245522e+01  +.30433822127772071943e+02
+.69729147895113773667e+00  +.17386456246734130171e+02
+.46483440633541432225e-01  +.20508906810000046669e+01
  84 6
+.36875639473458099862e+04  +.24806720777508070354e+01
+.25699663742660035012e+03  +.91185410850077186967e+01
+.32885250410274436152e+02  +.20141169412875392928e+02
+.55372807516587905839e+01  +.30533455818589890446e+02
+.76695240190109891076e+00  +.17863120770146898394e+02
+.77909417059715903711e-01  +.28630408356292925010e+01
  85 6
+.38672700297383376974e+04  +.24538110315323277185e+01
+.27079708409449999366e+03  +.89854236795571566622e+01
+.34834221835135341239e+02  +.19932186171355124066e+02
+.59163326065550116963e+01  +.30613532620204298900e+02
+.83714257744570818238e+00  +.18373965488949320224e+02
+.10024708223935701640e+00  +.36410810084017724287e+01
  86 6
+.40511370697500908635e+04  +.24300095420946064498e+01
+.28477593787039301706e+03  +.88632447318808189517e+01
+.36824980911702692865e+02  +.19727281480571014606e+02
+.63096008469553255530e+01  +.30677132568876660600e+02
+.90780562491788329230e+00  +.18918099140596643612e+02
+.11834115297115231510e+00  +.43842325359802557809e+01
  87 6
+.43539847943817889650e+04  +.23429209770104878628e+01
+.31502218975440361631e+03  +.84217316762871401933e+01
+.41780895975171042604e+02  +.18761468929404893311e+02
+.73246351832093548642e+01  +.30084189577167919949e+02
+.12167680803927787948e+01  +.18988459624861476385e+02
+.18531646855260040879e+00  +.74012292152680822992e+01
  88 6
+.44483506218660839615e+04  +.23809454644266776589e+01
+.31637235237572965497e+03  +.85759331031683028743e+01
+.41599655197053687089e+02  +.19127747791660405618e+02
+.73320030188586031866e+01  +.30322680706875483984e+02
+.11859426306615156698e+01  +.19296658373611664160e+02
+.16058344554826791978e+00  +.72960345602574657047e+01
  89 6
+.46031531564153656986e+04  +.23903650383882533962e+01
+.32430071106692241723e+03  +.86119611299271118442e+01
+.42437246625992947596e+02  +.19293262342063462455e+02
+.74892450697690965890e+01  +.30584541746206084794e+02
+.11874081908615181964e+01  +.19606439682466565325e+02
+.15164436711332391191e+00  +.75134300609485221858e+01
  90 6
+.47971151030446662116e+04  +.23795898439698880478e+01
+.33826880192816451082e+03  +.85331468200544571276e+01
+.44295335407383541629e+02  +.19245716363499046028e+02
+.78319083873759350076e+01  +.30679118109521893171e+02
+.12477414312105594584e+01  +.20130670917808011706e+02
+.16327011497437024786e+00  +.80317579451467039196e+01
  91 6
+.49883843473347966388e+04  +.23753063722205039939e+01
+.35079261962815327693e+03  +.84911083041889169853e+01
+.45860626299108343141e+02  +.19275177917826605551e+02
+.81275157307497156059e+01  +.30760434782802710068e+02
+.12948293491755238100e+01  +.20707670300884711543e+02
+.17433619558418046722e+00  +.83903023220765518582e+01
  92 6
+.51830661629723292921e+04  +.23735771858370988371e+01
+.36286617208313892515e+03  +.84648705410982652496e+01
+.47296628009567850930e+02  +.19347923356133982310e+02
+.83931120859276319413e+01  +.30855948029935142422e+02
+.13330953102037347731e+01  +.21320965705131993097e+02
+.18401729827822357817e+00  +.86367151818635180847e+01
  93 6
+.53845435728636195778e+04  +.23728653367085984718e+01
+.37492172838433873261e+03  +.84456432596417946307e+01
+.48682558935056744437e+02  +.19443484432954414085e+02
+.86430481254931232728e+01  +.30959928495385360443e+02
+.13667738267877879989e+01  +.21974025687078247004e+02
+.19260759242970482328e+00  +.88040527882315853653e+01
  94 6
+.55934551282203818809e+04  +.23726278910298545099e+01
+.38713773741553118690e+03  +.84296469743125632458e+01
+.50053950854638520089e+02  +.19552348078095362070e+02
+.88839185912501918286e+01  +.31069479019458111938e+02
+.13980540898103982687e+01  +.22664107845809891494e+02
+.20029695125356591674e+00  +.89117901912942167435e+01
  95 6
+.58115067109913875566e+04  +.23725743252191013815e+01
+.39961334346660233838e+03  +.84152488175778178302e+01
+.51428296016358131932e+02  +.19669688985931146693e+02
+.91189817697641509847e+01  +.31183102402744375203e+02
+.14281829164826449570e+01  +.23386365657937360861e+02
+.20721869574024736989e+00  +.89730198105901980308e+01
  96 6
+.60386361359839480956e+04  +.23726569093530474304e+01
+.41238355871728137540e+03  +.84016889149116913230e+01
+.52813831740458982918e+02  +.19792887877443171323e+02
+.93498288951217797018e+01  +.31300100663656808533e+02
+.14578869343677483154e+01  +.24135610134907145485e+02
+.21346404640287796287e+00  +.89970554997281359065e+01
  97 6
+.62609488163902367501e+04  +.23796302236646932081e+01
+.42291662308239006994e+03  +.84243404418558817247e+01
+.53904612510563164350e+02  +.19928274109709609692e+02
+.96025705329941451607e+01  +.31226916431527836086e+02
+.14984353150383831295e+01  +.25050862462418226131e+02
+.21896757528599816835e+00  +.89899763308237531581e+01
  98 6
+.64897401346844981036e+04  +.23876765106037633797e+01
+.43334955871057994186e+03  +.84509128186708503733e+01
+.54944806456545325941e+02  +.20084624843463702932e+02
+.98373295552491495576e+01  +.31163861867671763933e+02
+.15380653646173516523e+01  +.25941496882216141721e+02
+.22430150549872171494e+00  +.89714270773737776611e+01
  99 6
+.67261549694950843098e+04  +.23966473209503814847e+01
+.44372044537712748936e+03  +.84806183819026308722e+01
+.55944046178151569608e+02  +.20258712950977062225e+02
+.10056118549065121330e+02  +.31108117561816194970e+02
+.15773227998811425642e+01  +.26811850632963028763e+02
+.22947752376909677215e+00  +.89440531513907016842e+01
 100 6
+.69701361898739727637e+04  +.24064824188263240747e+01
+.45404343187592335530e+03  +.85128850639260001579e+01
+.56908789981374727584e+02  +.20448351921164018295e+02
+.10259912457405378162e+02  +.31058804932616764915e+02
+.16164506247480043634e+01  +.27664220086851843995e+02
+.23449400578056605322e+00  +.89092555766150485618e+01
 101 6
+.72231442425609014323e+04  +.24170541416903781674e+01
+.46434828194783742001e+03  +.85472803929013371146e+01
+.57845026820999606173e+02  +.20651830520976053376e+02
+.10449227089089557521e+02  +.31016261177045224748e+02
+.16555508524664383349e+01  +.28499613539628402704e+02
+.23935006172299832579e+00  +.88679602277586038902e+01
 102 6
+.74856133966236282386e+04  +.24282610034738251006e+01
+.47466477442887573459e+03  +.85832214235225030873e+01
+.58760145327839364080e+02  +.20867214864905726500e+02
+.10624698106362063148e+02  +.30981660197343698257e+02
+.16946873774494313422e+01  +.29318647910565503535e+02
+.24404577984956426995e+00  +.88209946001887435202e+01
\end{verbatim}


\begin{thebibliography}{10}
\providecommand{\url}[1]{{#1}}
\providecommand{\urlprefix}{URL }
\expandafter\ifx\csname urlstyle\endcsname\relax
  \providecommand{\doi}[1]{DOI \discretionary{}{}{}#1}\else
  \providecommand{\doi}{DOI \discretionary{}{}{}\begingroup
  \urlstyle{rm}\Url}\fi

\bibitem{H1963}
R.~Hoffmann, J. Chem. Phys. \textbf{39}, 1397 (1963).
\newblock \doi{10.1063/1.1734456}

\bibitem{KSKWP1967}
H.F. King, R.E. Stanton, H.~Kim, R.E. Wyatt, R.G. Parr, J. Chem. Phys.
  \textbf{47}, 1936 (1967).
\newblock \doi{10.1063/1.1712221}

\bibitem{SBBEGJKMNSWDM1993}
M.W. Schmidt, K.K. Baldridge, J.A. Boatz, S.T. Elbert, M.S. Gordon, J.H.
  Jensen, S.~Koseki, N.~Matsunaga, K.A. Nguyen, S.~Su, T.L. Windus, M.~Dupuis,
  J.A. Montgomery, J. Comput. Chem. \textbf{14}, 1347 (1993).
\newblock \doi{10.1002/jcc.540141112}

\bibitem{HAKRST1984}
S.~Huzinaga, J.~Andzelm, M.~Klobukowski, E.~Radzio-Andzelm, Y.~Sakai,
  H.~Tatewaki, \emph{Gaussian basis sets for molecular calculations},
  \emph{Physical sciences data}, vol.~16 (Elsevier, Amsterdam, 1984)

\bibitem{SH1953}
R.T. Sharp, G.K. Horton, Phys. Rev. \textbf{90}, 317 (1953).
\newblock \doi{10.1103/physrev.90.317}

\bibitem{TS1976}
J.D. Talman, W.F. Shadwick, Phys. Rev. A \textbf{14}, 36 (1976).
\newblock \doi{10.1103/physreva.14.36}

\bibitem{MSBG2011}
P.~Maldonado, A.~Sarsa, E.~Buend{\'{\i}}a, F.~G{\'{a}}lvez, Atom. Data Nucl.
  Data \textbf{97}, 109 (2011).
\newblock \doi{10.1016/j.adt.2010.10.002}

\bibitem{KS1965}
W.~Kohn, L.J. Sham, Phys. Rev. \textbf{140}, A1133 (1965).
\newblock \doi{10.1103/PhysRev.140.A1133}

\bibitem{H1985}
J.~Harris, Phys. Rev. B \textbf{31}, 1770 (1985).
\newblock \doi{10.1103/physrevb.31.1770}

\bibitem{B1988}
A.D. Becke, J. Chem. Phys. \textbf{88}, 2547 (1988).
\newblock \doi{10.1063/1.454033}

\bibitem{LKO2018}
H.~Laqua, J.~Kussmann, C.~Ochsenfeld, J. Chem. Phys. \textbf{149}, 204111
  (2018).
\newblock \doi{10.1063/1.5049435}

\bibitem{Lehtola2019}
S.~Lehtola, J. Chem. Theory Comput. \textbf{15}, 1593 (2019).
\newblock \doi{10.1021/acs.jctc.8b01089}

\bibitem{LZDG2006}
J.H. Van~Lenthe, R.~Zwaans, H.J.J. Van~Dam, M.F. Guest, J. Comput. Chem.
  \textbf{27}, 926 (2006).
\newblock \doi{10.1002/jcc.20393}

\bibitem{AC2001}
L.~Amat, R.~Carb{\'{o}}-Dorca, Int. J. Quantum Chem. \textbf{87}, 59 (2001).
\newblock \doi{10.1002/qua.10068}

\bibitem{SF1975}
H.~Sambe, R.H. Felton, J. Chem. Phys. \textbf{62}, 1122 (1975).
\newblock \doi{10.1063/1.430555}

\bibitem{DCS1979}
B.I. Dunlap, J.W.D. Connolly, J.R. Sabin, J. Chem. Phys. \textbf{71}, 3396
  (1979).
\newblock \doi{10.1063/1.438728}

\bibitem{S1951}
J.C. Slater, Phys. Rev. \textbf{81}, 385 (1951).
\newblock \doi{10.1103/PhysRev.81.385}

\bibitem{D1986}
B.I. Dunlap, J. Phys. Chem. \textbf{90}, 5524 (1986).
\newblock \doi{10.1021/j100280a010}

\bibitem{L1997}
D.N. Laikov, Chem. Phys. Lett. \textbf{281}, 151 (1997).
\newblock \doi{10.1016/S0009-2614(97)01206-2}

\bibitem{NW2017}
F.~Nazari, J.L. Whitten, J. Chem. Phys. \textbf{146}, 194109 (2017).
\newblock \doi{10.1063/1.4983395}

\bibitem{W2019}
J.L. Whitten, J. Chem. Phys. \textbf{150}, 034107 (2019).
\newblock \doi{10.1063/1.5064781}

\bibitem{W2019b}
J.L. Whitten, Phys. Chem. Chem. Phys. \textbf{21}, 21541 (2019).
\newblock \doi{10.1039/c9cp02450f}

\bibitem{B1950}
S.F. Boys, Proc. R. Soc. A \textbf{200}, 542 (1950).
\newblock \doi{10.1098/rspa.1950.0036}

\bibitem{L2019b}
D.N. Laikov, Theor. Chem. Acc. \textbf{138}, 40 (2019).
\newblock \doi{10.1007/s00214-019-2432-3}

\bibitem{D1994}
K.G. Dyall, J. Chem. Phys. \textbf{100}, 2118 (1994).
\newblock \doi{10.1063/1.466508}

\bibitem{VD1997}
L.~Visscher, K.G. Dyall, Atom. Data Nucl. Data \textbf{67}, 207 (1997).
\newblock \doi{10.1006/adnd.1997.0751}

\bibitem{L2019a}
D.N. Laikov, J. Chem. Phys. \textbf{150}, 061103 (2019).
\newblock \doi{10.1063/1.5082231}

\bibitem{KBT76}
L.R. Kahn, P.~Baybutt, D.G. Truhlar, J. Chem. Phys. \textbf{65}, 3826 (1976).
\newblock \doi{10.1063/1.432900}

\bibitem{PBE1996}
J.P. Perdew, K.~Burke, M.~Ernzerhof, Phys. Rev. Lett. \textbf{77}, 3865 (1996).
\newblock \doi{10.1103/PhysRevLett.77.3865}

\bibitem{L2005}
D.N. Laikov, Chem. Phys. Lett. \textbf{416}, 116 (2005).
\newblock \doi{10.1016/j.cplett.2005.09.046}

\bibitem{L2011}
D.N. Laikov, J. Chem. Phys. \textbf{135}, 134120 (2011).
\newblock \doi{10.1063/1.3646498}

\bibitem{MD1978}
L.E. McMurchie, E.R. Davidson, J. Comput. Phys. \textbf{26}, 218 (1978).
\newblock \doi{10.1016/0021-9991(78)90092-x}

\bibitem{HK1983}
S.~Havriliak, H.F. King, J. Am. Chem. Soc. \textbf{105}, 4 (1983).
\newblock \doi{10.1021/ja00339a002}

\bibitem{WP2003}
K.~Wolinski, P.~Pulay, J. Chem. Phys. \textbf{118}, 9497 (2003).
\newblock \doi{10.1063/1.1562606}

\bibitem{DGG2009}
J.~Deng, A.T.B. Gilbert, P.M.W. Gill, J. Chem. Phys. \textbf{130}, 231101
  (2009).
\newblock \doi{10.1063/1.3152864}

\bibitem{DGG2010}
J.~Deng, A.T.B. Gilbert, P.M.W. Gill, J. Chem. Phys. \textbf{133}, 044116
  (2010).
\newblock \doi{10.1063/1.3463800}

\bibitem{MBP2009}
J.~Martin, J.~Baker, P.~Pulay, J. Comput. Chem. \textbf{30}, 881 (2009).
\newblock \doi{10.1002/jcc.21106}

\end{thebibliography}
\end{document}